**Short Technical Report**: Best student-led project in COMP0145: Research Methods and Making Skills (2022/23)

# Self-Assistant: Portable Progressive Muscle Relaxation Training Interface for Anxiety Reduction in Office Workers

Lingqian Yang, Katherine Wang, and Youngjun Cho*

University College London

Abstract: Workload often triggers anxiety for office workers. While a variety of stress intervention and management techniques have been explored, there exist only a few of portable tangible interfaces for anxiety reduction. Contributing to the body of work, we introduce Self-Assistant, a portable anxiety intervention training interface. This is based on progressive muscle relaxation training which is an effective way for reducing stress and raise awareness of tension, and provide feelings of deep relaxation. This interface incorporates a stress ball with a pressure sensor and visual indicators which help track the user's anxiety level through squeeze actions. This is designed to help improve self-awareness of anxiety and stress, in turn self-regulating bodily functions, offering rapid and low-sensory stimulus training for anxiety relief. Finally, we discuss application scenarios and future work with Self-Assistant.

Keywords: *Anxiety reduction, Progressive Muscle Relaxation, Self-care intervention*

## 1 INTRODUCTION

Stress and anxiety tremendously affect office workers' mental health [1,2]. Research has shown that office workers experience stressful situations more frequently than other groups who tend to be physically active (e.g., manufacturing sectors) due to the sedentary nature of work [3]. Stress is a major risk factor for physical health conditions as well, such as high blood pressure, cardiovascular disease, gastrointestinal problems, headaches, and back pain [6,7]. People may experience a noticeable strong, fast or irregular heartbeat and muscle aches and tensions, trembling or shaking, or in more severe cases, anxiety attacks. Given that average working hours of office workers account for a large portion of daily life (e.g., over 36hours per week in Europe [4]), it is important to proactively manage daily stress. Alleviating this can also dramatically improve their productivity and performance and promote physical and mental health [7].

A wide range of workplace stress interventions and management techniques have been studied [5,6,8]. These mainly include muscle relaxation, cognitive behavioral techniques, biofeedback and meditation, psychotropics, psychotherapy [5,6,8]. Among physical interventions, Progressive Muscle Relaxation (PMR) has been shown to be an effective way of reducing stress and raise awareness of tension and provide feelings of deep relaxation [8,9,10]. PMR has been used in anxiety treatment applications for people with tension headaches, migraines and insomnia [9,11]. Research suggests that multiple forms of PMR (e.g., progressive muscle relaxation, meditation, breathing exercises, visualization, and autogenic) can benefit employees by reducing work-related anxiety, enhance relaxation states, and improve overall well-being [9,10,12].

Simple PMR digital tutorials are available on the internet in text, audio, and video form, a few of them use physical forms like magnetic tape playing prerecorded audios [14]. It is suggested that practice should be done in a quiet place alone without distractions [12]. However, there are few physical, tangible and interactive solutions associated with such relaxation training where interactivity of, and motivation for, training is key to successful PMR and its effectiveness. Most studies focus on monitoring mental stress through biofeedback quantification by using wearable physiological sensing channels such as electrocardiogram (ECG) and photoplethysmography (PPG) [13,15,17,18].

Here, we introduce Self-Assistant, a portable and physical PMR interface for alleviating anxiety in office workers. This interactive device is designed to be easy-prompted with a simple visual-cued PMR training. This includes a foam ball that helps to track the user's anxiety level by reading the accumulated value from the embedded pressure sensor in it. PMR training phases are transitioned along with preset pressure values, with an LCD screen providing step-by-step text instructions and an LED for a silent countdown with its connected mobile app. This builds on the literature review of existing studies on stress balls designed for anxiety regulation.

## 2 RELATED WORK

The use of tactile devices can be useful for alleviating stress [16,19,20]. As a manipulative interaction technique with tactile feedback, squeeze action is considered to be good at improving performance [23]. For instance, stress balls made of soft materials are widely used to release built-up frustration when squeezed repeatedly. Stress balls implemented with different interactive functions have been explored as anxiety interventions [24,25,27]. For instance, some have been implemented with sensors that retrieve and quantify pressure levels [21,22,26]. Most research involving these stress balls focuses on reading, visualization, and feedback on anxiety data [19,21,22]. For example, Clasp [21] connects a tactile anxiety coping device to a smartphone to record and communicate anxiety levels for self-feedback and reflection. Also, SWARM [22], a wearable emotion technology uses a scarf embedded with stress balls to support expression and transmission of emotions and to provide emotional feedback.

There is also compelling evidence to support the use of pressure-sensing tactile devices in anxiety interventions for different types of populations. For instance, autistic children have been shown to build a strong attachment to soft toys embedded with a pressure ball and demonstrate increased levels of concentration [16]. However, there is still a gap in these interventions in incorporating PMR training into the tangible interactive interfaces for active anxiety intervention. In this report, we focus on designing a self-intervention interface for adults which triggers the process of PMR treatment with a pressure-reading stress ball connected to an interactive mobile app.

## 3 PROPOSED INTERFACE: SELF-ASSISTANT

Managing anxiety is crucial to ensuring comfortable working conditions for employees and contributes to productivity. Based upon the literature of PMR training and anxiety regulation techniques, self-intervention with a tangible interface is proved to be an efficient solution. The specific design goals are as follows:

- Integrate the presence of the stress ball into office contexts, allowing the user to form the habit of squeezing the ball to self-record anxiety levels and their historical changes along with its connected mobile app (Self-Assistant in Figure 1).
- Visualize of the user's anxiety level with comfortable illuminated lights based on the squeeze action (Figure 2), enhancing the user's awareness of their emotional state change.
- Provide explicit and simple step-by-step guidelines on the PMR training in a personal and silent environment (Figure 3, 4).
- Set silent mode and demonstrate it by subtle blue lighting to convey the users' desire to not be disturbed.

The Self-Assistant is a small portable tactile device with an app which tracks user anxiety levels and guides users through the Progressive Muscle Relaxation training by providing straightforward visual instructions.

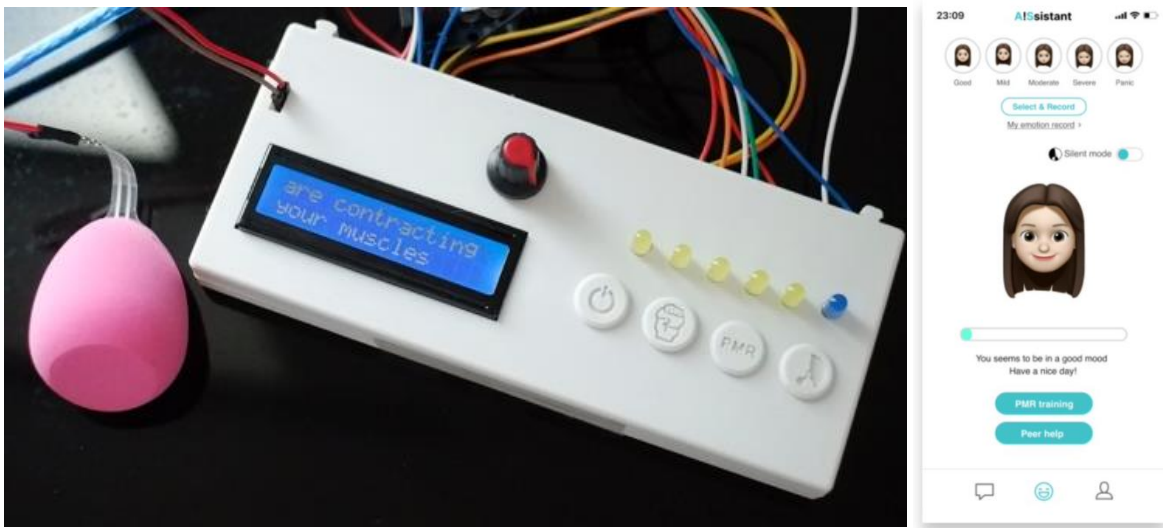

Figure 1. Proposed Progressive Muscle Relaxation Training Interface for Anxiety Reduction: Self-Assistant (left: tangible interface; right: connected mobile app)

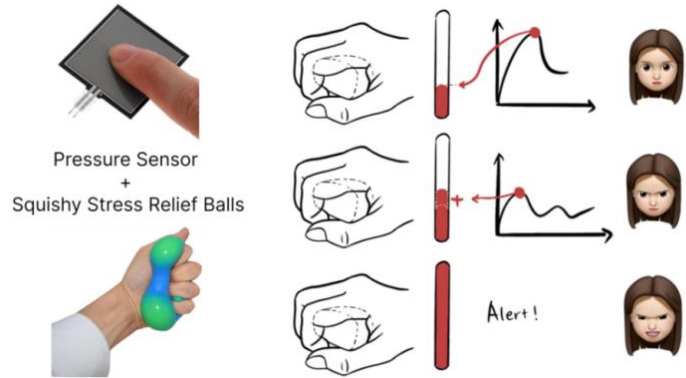

Figure 2. Mechanism for recording the user's anxiety level (user input through pressure sensing).

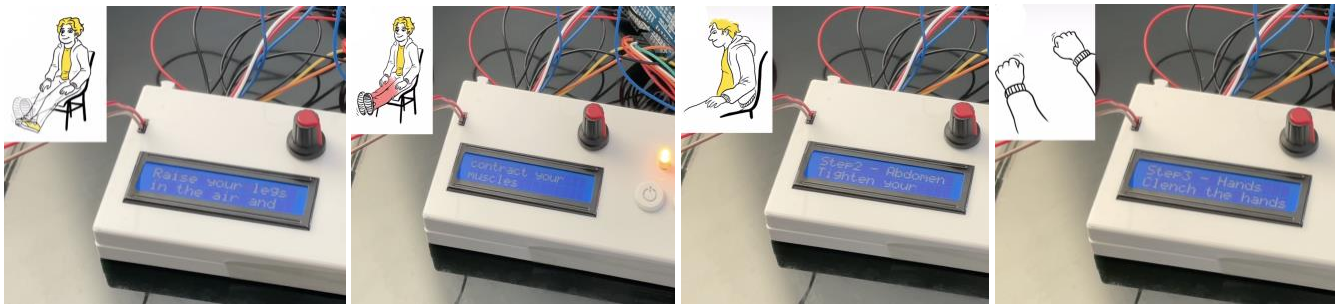

Figure 3. Explicit and simple step-by-step guidelines for PMR training.

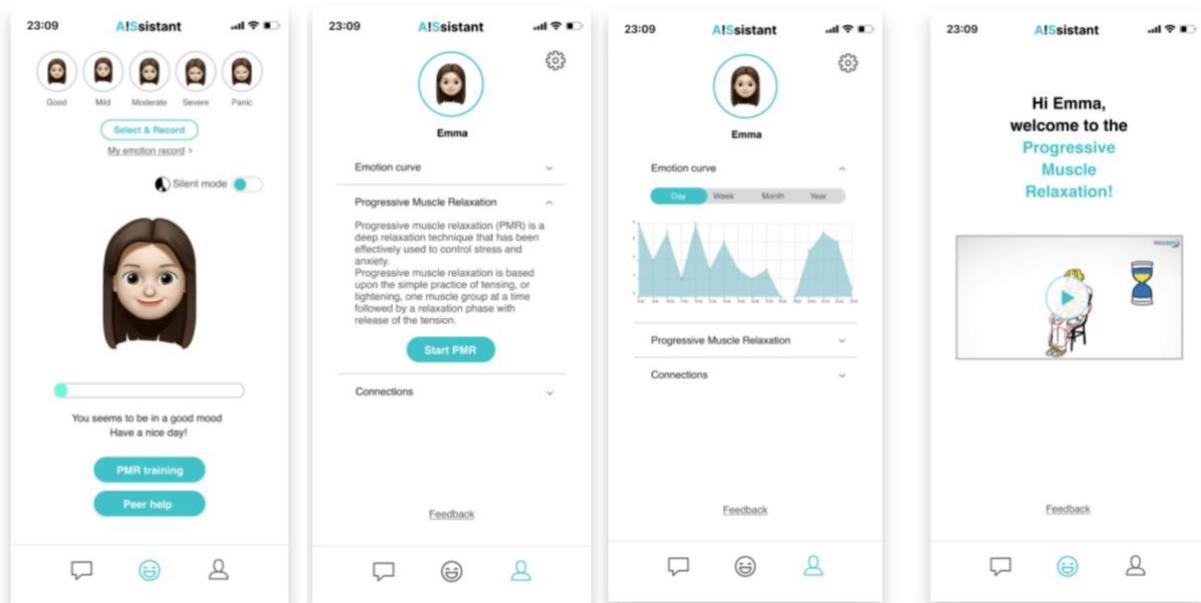

Figure 4. Self-Assistant mobile app interface that effectively visualize users' anxiety level changes over time and prompt the PMR-based active anxiety regulation.

If no other buttons are triggered, the device will be ready to detect users' anxiety level (user input through pressure sensing as shown in Figure 2) by default. As the user squeezes the sponge ball, the anxiety level will be recorded and accumulated gradually (Figure 2) and will be indicated through the gradual illumination of the LED. When the pressure level hits its maximum, the LED lights are all lit. Then the device will show a line on the LCD display to suggest that the user should start PMR training with its connected mobile app (Figure 3 and 4). For the PMR training, each step will be counted and indicated with a row of LEDs changing their brightness levels. We use bright lights which are less distracting and obvious enough to gain attention. With the silent mode button (with the blue light), users can inform others of anxiety states so as not to be disturbed. Further details on the user journey can be found in Appendix.

**4 FUTURE WORK**

The Self-Assistant is based on a set of soothing PMR training and tracking methods that have been proven to be effective. This interface can be of help for users in workplaces as well as remote working environments where users can easily access the training and perform self-intervention. In our future work, we plan to conduct a study on the Self-Assistant intervention for anxiety reduction, particularly understanding how this can benefit office workers. It would also be interesting to explore how active tactile feedback can be integrated to make it more interactive [28,29]. Further, we will explore how physiological sensing can help personalize the PMR training and improve the effectiveness [30,31,32]. Lastly, it would be an interesting direction to create a private channel for facilitating emotional communication between colleagues. This can possibly be alternative forms of workplace informal communication to make people to feel more connected.

**REFERENCES**

[1] Laura Goodwin, Ilan Ben-Zion, Nicola T. Fear, Matthew Hotopf, Stephen A. Stansfeld, and Simon Wessely. 2013. Are Reports of Psychological Stress Higher in Occupational Studies? A Systematic Review across Occupational and Population Based Studies. PLoS ONE 8, 11 (November 2013), e78693. DOI:https://doi.org/10.1371/journal.pone.0078693

[2] Stephen Alfred Stansfeld, F. R. Rasul, J. Head, and N. Singleton. 2009. Occupation and mental health in a national UK survey. Social Psychiatry and Psychiatric Epidemiology 46, 2 (December 2009), 101–110. DOI:https://doi.org/10.1007/s00127-009-0173-7

[3] WonYang Kang, Won-Ju Park, Keun-Ho Jang, Hyeong-Min Lim, Ji-Sung Ann, Seung-hyeon Cho, and Jai-Dong Moon. 2016. Comparison of anxiety and depression status between office and manufacturing job employees in a large manufacturing company: a cross sectional study. Annals of Occupational and Environmental Medicine 28, 1 (September 2016). DOI:https://doi.org/10.1186/s40557-016-0134-z

[4] Hours of work - annual statistics. ec.europa.eu. Retrieved January 17, 2023 from https://ec.europa.eu/eurostat/statistics-explained/index.php?title=Hours_of_work_-_annual_statistics#:~:text=Highlights&text=In%202021%2C%20the%20average%20working

[5] Katherine M. Richardson and Hannah R. Rothstein. 2008. Effects of occupational stress management intervention programs: A meta-analysis. Journal of Occupational Health Psychology 13, 1 (2008), 69–93. DOI:https://doi.org/10.1037/1076-8998.13.1.69

[6] Lawrence R. Murphy. 1996. Stress Management in Work Settings: A Critical Review of the Health Effects. American Journal of Health Promotion 11, 2 (November 1996), 112–135. DOI:https://doi.org/10.4278/0890-1171-11.2.112

[7] Margaret Vaughn, Susan Cheatwood, Ann T. Sirles, and Kathleen C. Brown. 1989. The Effect of Progressive Muscle Relaxation on Stress among Clerical Workers. AAOHN Journal 37, 8 (August 1989), 302–306. DOI:https://doi.org/10.1177/216507998903700801

[8] Julio Torales, Marcelo O'Higgins, and Israel Gonzalez. 2020. An Overview of Jacobson's Progressive Muscle Relaxation in Managing Anxiety. (May 2020). DOI:https://doi.org/10.24205/03276716.2020.748

[9] Loren Toussaint, Quang Anh Nguyen, Claire Roettger, Kiara Dixon, Martin Offenbächer, Niko Kohls, Jameson Hirsch, and Fuschia Sirois. 2021. Effectiveness of Progressive Muscle Relaxation, Deep Breathing, and Guided Imagery in Promoting Psychological and Physiological States of Relaxation. Evidence-Based Complementary and Alternative Medicine. Retrieved from https://www.hindawi.com/journals/ecam/2021/5924040/

[10] Ahmad Fairuz Mohamed, Marzuki Isahak, Mohd Zaki Awg Isa, and Rusli Nordin. 2022. The effectiveness of workplace health promotion program in reducing work-related depression, anxiety and stress among manufacturing workers in Malaysia: mixed-model intervention. International Archives of Occupational and Environmental Health (January 2022). DOI:https://doi.org/10.1007/s00420-022-01836-w

[11] Kai Liu, Ying Chen, Duozhi Wu, Ruzheng Lin, Zaisheng Wang, and Liqing Pan. 2020. Effects of progressive muscle relaxation on anxiety and sleep quality in patients with COVID-19. Complementary Therapies in Clinical Practice 39, (May 2020), 101132. DOI:https://doi.org/10.1016/j.ctcp.2020.101132

[12] Martha S. McCallie, Claire M. Blum, and Charlaine J. Hood. 2006. Progressive Muscle Relaxation. Journal of Human Behavior in the Social Environment 13, 3 (July 2006), 51–66. DOI:https://doi.org/10.1300/j137v13n03_04

[13] Božidara Cvetković, Martin Gjoreski, Jure Šorn, Pavel Maslov, Michał Kosiedowski, Maciej Bogdański, Aleksander Stroiński, and Mitja Luštrek. 2017. Real-time physical activity and mental stress management with a wristband and a smartphone. Proceedings of the 2017 ACM International Joint Conference on Pervasive and Ubiquitous Computing and Proceedings of the 2017 ACM International Symposium on Wearable Computers (September 2017). DOI:https://doi.org/10.1145/3123024.3123184

[14] Sermsak Lolak, Gerilynn L. Connors, Michael J. Sheridan, and Thomas N. Wise. 2008. Effects of Progressive Muscle Relaxation Training on Anxiety and Depression in Patients Enrolled in an Outpatient Pulmonary Rehabilitation Program. Psychotherapy and Psychosomatics 77, 2 (2008), 119–125. DOI:https://doi.org/10.1159/000112889

[15] Youngjun Cho, Simon J. Julier, and Nadia Bianchi-Berthouze. 2019. Instant Stress: Detection of Perceived Mental Stress Through Smartphone Photoplethysmography and Thermal Imaging. JMIR Mental Health 6, 4: e10140. https://doi.org/10.2196/10140

[16] Will Simm, Maria Ferrario, Jen Southern, and Jon Whittle. PAPER: Clasp: Digital Tactile ASD Anxiety Management. Retrieved from https://www.lancaster.ac.uk/myclasp/wp-content/uploads/2015/03/DE2013_Clasp.pdf

[17] Clara Moge, Katherine Wang, and Youngjun Cho. 2022. Shared User Interfaces of Physiological Data: Systematic Review of Social Biofeedback Systems and Contexts in HCI. In Proceedings of the 2022 CHI Conference on Human Factors in Computing Systems (CHI '22), 1–16.

https://doi.org/10.1145/3491102.3517495

[18] Youngjun Cho, Sunuk Kim, and Munchae Joung. 2017. Proximity Sensor and Control Method Thereof. US Patent, patent number: 9703368.

[19] Jinsil Hwaryoung Seo, Pavithra Aravindan, and Annie Sungkajun. 2017. Toward Creative Engagement of Soft Haptic Toys with Children with Autism Spectrum Disorder. Proceedings of the 2017 ACM SIGCHI Conference on Creativity and Cognition (June 2017). DOI:https://doi.org/10.1145/3059454.3059474

[20] R. Srivarsan, G. Sridevi, and S. Preetha. 2021. An Evaluation on Use of Stress Ball Exercise on Stress Management among Student Population – A Cross Section Study. Journal of Pharmaceutical Research International (November 2021), 506–514. DOI:https://doi.org/10.9734/jpri/2021/v33i47b33150

[21] Will Simm, Maria Angela Ferrario, Adrian Gradinar, and Jon Whittle. 2014. Prototyping 'clasp': implications for designing digital technology for and with adults with autism. Proceedings of the 2014 conference on Designing interactive systems (June 2014). DOI:https://doi.org/10.1145/2598510.2600880

[22] Michele A. Williams, Asta Roseway, Chris O'Dowd, Mary Czerwinski, and Meredith Ringel Morris. 2015. SWARM. Proceedings of the Ninth International Conference on Tangible, Embedded, and Embodied Interaction (January 2015). DOI:https://doi.org/10.1145/2677199.2680565

[23] Eve Hoggan, Dari Trendafilov, Teemu Ahmaniemi, and Roope Raisamo. 2011. Squeeze vs. tilt. CHI '11 Extended Abstracts on Human Factors in Computing Systems (May 2011). DOI:https://doi.org/10.1145/1979742.1979766

[24] Fereshteh Shahmiri, Steven Schwartz, and Can Usanmaz. 2022. A Therapeutic Stress Ball to Monitor Hand Dexterity and Electrodermal Activity. arXiv:2202.12947 [physics] (February 2022). Retrieved December 19, 2022 from https://arxiv.org/abs/2202.12947

[25] Kadriye Sayin Kasar, Saadet Erzincanli, and Nesat Tolga Akbas. 2020. The effect of a stress ball on stress, vital signs and patient comfort in hemodialysis patients: A randomized controlled trial. Complementary Therapies in Clinical Practice 41, (November 2020), 101243. DOI:https://doi.org/10.1016/j.ctcp.2020.101243

[26] Ali Karime, A S M Mahfujur Rahman, Abdulmotaleb El Saddik, and Wail Gueaieb. 2011. RehaBall: Rehabilitation of upper limbs with a sensory-integrated stress ball. 2011 IEEE International Workshop on Haptic Audio Visual Environments and Games (October 2011). DOI:https://doi.org/10.1109/have.2011.6088388

[27] R. Srivarsan, G. Sridevi, and S. Preetha. 2021. An Evaluation on Use of Stress Ball Exercise on Stress Management among Student Population – A Cross Section Study. apsciencelibrary.com (November 2021). Retrieved from http://apsciencelibrary.com/handle/123456789/6587

[28] Dong-Soo Kwon, Tae-Heon Yang, and Young-Jun Cho. 2010. Mechatronics Technology in Mobile Devices. IEEE Industrial Electronics Magazine 4, 2: 36–41. https://doi.org/10.1109/MIE.2010.936763

[29] Anastasia Schmitz, Catherine Holloway, and Youngjun Cho. 2020. Hearing through Vibrations: Perception of Musical Emotions by Profoundly Deaf People. https://doi.org/10.48550/arXiv.2012.13265

[30] Youngjun Cho, Nadia Bianchi-Berthouze, Simon J. Julier, and Nicolai Marquardt. 2017. ThermSense: Smartphone-based breathing sensing platform using noncontact low-cost thermal camera. In 2017 Seventh International Conference on Affective Computing and Intelligent Interaction Workshops and Demos (ACIIW), 83–84. https://doi.org/10.1109/ACIIW.2017.8272593

[31] Jitesh Joshi, Nadia Berthouze, and Youngjun Cho. 2022. Self-adversarial Multi-scale Contrastive Learning for Semantic Segmentation of Thermal Facial Images. In The 33rd British Machine Vision Conference Proceedings, 1–18. https://bmvc2022.mpi-inf.mpg.de/864/

[32] Youngjun Cho. 2017. Automated mental stress recognition through mobile thermal imaging. In 2017 Seventh International Conference on Affective Computing and Intelligent Interaction (ACII), 596–600. https://doi.org/10.1109/ACII.2017.8273662

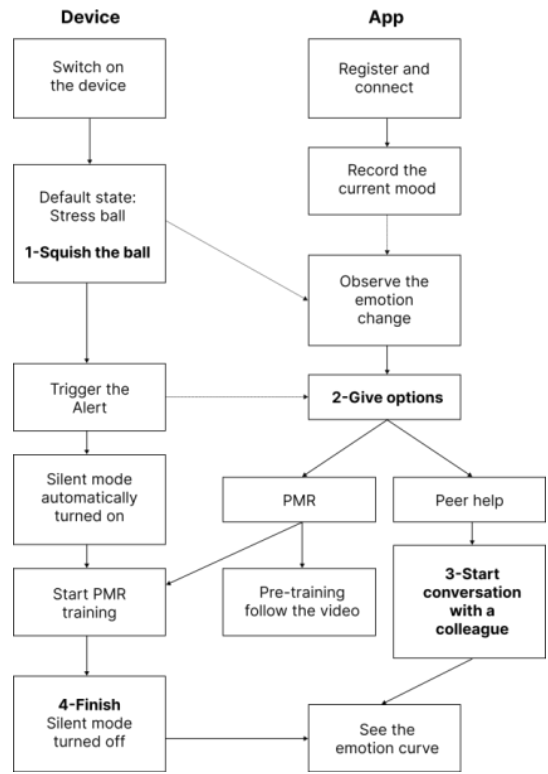

Appendix Figure: Further details on the user journey flow